\documentclass[prb,twocolumn,aps,superscriptaddress,showpacs]{revtex4-1}

\usepackage{amsmath,amssymb}
\usepackage{graphicx}
\usepackage{xcolor}

\begin{document}

\title{\textbf{Quantum Criticality in the Ferromagnetic Superconductor UCoGe under Pressure and Magnetic Field}}

\author{Ga\"{e}l Bastien}
\email{gael.bastien@cea.fr}
\affiliation{Universit\'{e} Grenoble Alpes, CEA, INAC-PHELIQS, F-38000 Grenoble, France}
\author{Daniel Braithwaite}
\affiliation{Universit\'{e} Grenoble Alpes, CEA, INAC-PHELIQS, F-38000 Grenoble, France}
\author{Dai Aoki}
\affiliation{Universit\'{e} Grenoble Alpes, CEA, INAC-PHELIQS, F-38000 Grenoble, France}
\affiliation{IMR, Tohoku University, Oarai, Ibaraki 311-1313, Japan}
\author{Georg Knebel}
\email{georg.knebel@cea.fr}
\affiliation{Universit\'{e} Grenoble Alpes, CEA, INAC-PHELIQS, F-38000 Grenoble, France}
\author{Jacques Flouquet}
\affiliation{Universit\'{e} Grenoble Alpes, CEA, INAC-PHELIQS, F-38000 Grenoble, France}

\date{\today }

\begin{abstract}
The pressure-temperature phase diagram of the orthorhombic ferromagnetic superconductor UCoGe was determined by resistivity measurements up to $10.5~$GPa. The Curie temperature $T_C$ is suppressed  with pressure and vanishes at the critical pressure $p_c\approx1~$GPa. Superconductivity is observed in both the ferromagnetic state at low pressure, and in the paramagnetic state above $p_c$ up to about 4$~$GPa. Non-Fermi liquid behavior appears in a large pressure range. The resistivity varies linearly with temperature around $p_c$ and evolves continuously with pressure to a $T^2$ Fermi-liquid behavior for $p \gtrapprox 5$~GPa. 
The residual resistivity as a function of pressure shows a maximum far above $p_c$ at $p^\star=7.2~$GPa and the amplitude of the inelastic scattering term of the resistivity decreases by more than one order in magnitude at $p^\star$, which appears to mark the entrance into a weakly correlated regime. The pressure dependence of the upper critical field for magnetic field applied along the $b$ and $c$ axis illustrates the drastic difference in the field dependence of the ferromagnetic superconducting pairing. While for $H\parallel b$ axis $H_{c2}(T)$ is driven by the suppression of the ferromagnetic order, it is dominated by the strong initial suppression of the ferromagnetic fluctuations for a field applied  in the easy magnetization axis $c$. 
\end{abstract}


\maketitle

\section{Introduction}

Magnetic quantum criticality has attracted major attention in recent years. The magnetic quantum phase transition is accompanied  by an increase of  quantum fluctuations, which induces strong deviations from the standard Fermi-liquid behavior of a metal  \cite{Loehneysen2007}. New quantum phases  such as magnetically mediated non-conventional superconductivity (SC) \cite{Pfleiderer2009}, nematic orders \cite{Borzi2007, Fernandes2014}, or complex modulated magnetic states \cite{Abdul-Jabbar2015} appear near the critical pressure $p_c$.  

The collapse of antiferromagnetism  occurs in many heavy fermion systems through a continuous second order quantum phase transition, and a superconducting dome has been observed in the vicinity of $p_c$. In contrast, in the case of a ferromagnet, the quantum phase transition at $p_c$  changes near $p_c$ to be first order in  systems with low disorder \cite{Belitz2005, Mineev2011} or may lead to complex order \cite{Chubukov2004, Conduit2009, Karahasanovic2012}. Even if magnetically mediated superconductivity has been predicted for weakly ferromagnetic systems already long time ago \cite{Fay1980}, up to now only four systems show the microscopic coexistence of ferromagnetism and SC, UGe$_2$ \cite{Saxena2000}, URhGe \cite{Aoki2001}, UIr \cite{Akazawa2004}, and UCoGe \cite{Huy2007}. 
While for UGe$_2$ SC appears under high pressure in the ferromagnetic state, URhGe and UCoGe are ferromagnetic and superconducting at already ambient pressure but with very different $(p,T)$ phase diagrams. In URhGe the Curie temperature $T_{C}$ increases with pressure at least up to 12.6~GPa \cite{Hardy2005} while ferromagnetism in UCoGe is suppressed under pressure \cite{Hassinger2008b, Slooten2009}. 

In this article we focus on the high pressure phase diagram of UCoGe which orders at $T_{C}\approx 2.7$~K with a tiny ordered moment of $M_0\approx 0.05~\mu_B$ at $p=0$ \cite{Huy2007}.  Ferromagnetism is due to the 5$f$ electrons from uranium which hybridize with light electrons to form heavy quasiparticle bands at low temperature \cite{Fujimori2015}.  Bulk superconductivity coexist with ferromagnetism below $T_{sc}\approx 0.6$~K.
This provides an unique opportunity to study the link between the collapse of ferromagnetism  and concomitant ferromagnetic fluctuations and the appearance of SC. Previously, the $(p,T)$ phase diagram of UCoGe has been established up to 2.5$~$GPa by resistivity and magnetic susceptibility measurements \cite{Hassinger2008b, Slooten2009, Hassinger2010a, Bastien2015} and $p_c$ is located around 1$~$GPa.  The ferromagnetic transition may be already  first order at ambient pressure \cite{Ohta2010}. 
Here we will determine the $(p,T)$ phase diagram of UCoGe by resistivity measurements up to 10.5~GPa down to 50~mK using a diamond anvil cell (DAC).
 Our experiments illustrate the link between  the normal and SC phases on the field and pressure dependence of the ferromagnetic coupling. UCoGe provides the unique opportunity to study the interplay between SC and ferromagnetism in a case where SC is known to be robust through $p_c$ and where $T_C$ will approach $T_{sc}$ without a strong first order collapse of $T_C$ as it happens for example in UGe$_2$.\cite{Pfleiderer2002}

In UCoGe SC survives  deep inside the paramagnetic (PM) domain roughly up to $p\approx 4 p_c$. Up to this pressure magnetic fluctuations are dominant and the transport shows a
non-Fermi-liquid regime in the normal state above the SC dome. A Fermi-liquid ground state in zero field can be detected only for $p \gtrapprox 5$~GPa, basically when SC has fully collapsed. The magnetic field and pressure variations of ferromagnetism and its concomitant fluctuations have strong impact on the  unusual temperature dependence of the superconducting upper critical field $H_{c2}$ with an upward curvature observed initially for both, the hard $b$ and the easy $c$ axis at low pressure \cite{Bastien2015, Aoki2011c}.

\section{Experimental Details}

High quality single crystals of UCoGe (orthorhombic TiNiSi structure) were grown in a tetra-arc furnace by the Chzochralski method. Measurements under hydrostatic pressure were performed in a DAC ($p< 11$~GPa) and a piston cylinder cell ($p< 2.5$~GPa) with argon and Daphne oil 7373  as pressure transmitting media, respectively, ensuring very good pressure conditions \cite{Tateiwa2009}. 
The resistivity ratio  of the sample used in the DAC after polishing was RRR = 28. A dilution fridge with a system allowing to change pressure in-situ at low temperature was used to cool the sample down to 50~mK  \cite{bellow, Salce2000}. The pressure was determined in-situ by fluorescence of ruby. An ac electrical current ($j < 100\;\mu$A) was applied in the $ab$ plane. A low temperature transformer has been used to improve the signal to noise ratio of the measured voltage. Magnetic field up to 7~T was applied along the $c$ axis. In a second experiment,  the upper critical field $H_{c2}$ for $H \parallel b$ has been determined by resistivity with  a sample of similar quality using a piston cylinder cell up to 2.5~GPa and magnetic field up to 16~T.

\section{Results and Discussion}

\begin{figure}[!t]
\begin{center}
\includegraphics[width=0.9\linewidth]{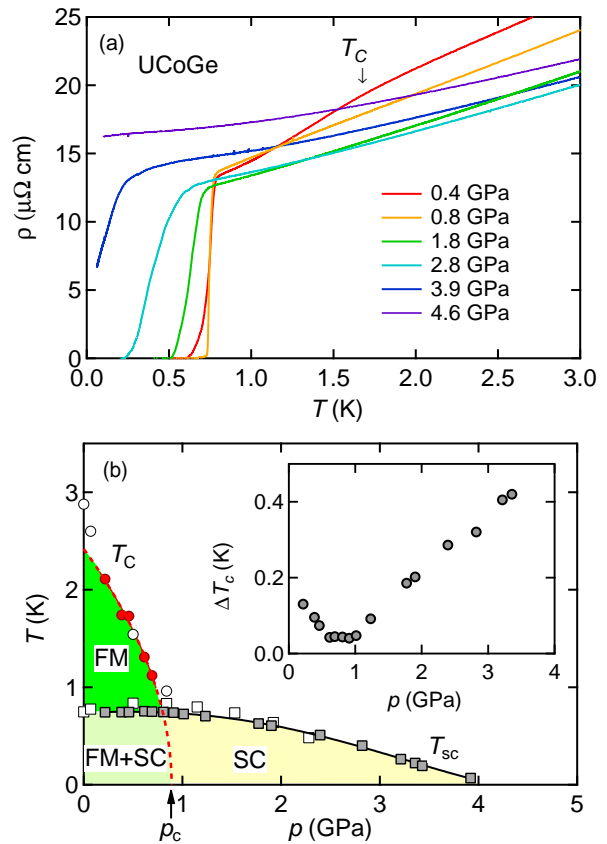}
\caption{(Color online) (a) Temperature dependence of the resistivity in UCoGe for different pressures in the diamond anvil cell. For $p=0.4$~GPa the Curie temperature $T_C$ is denoted by the arrow. (b) ($p,T$) phase diagram of UCoGe. Closed (open) symbols correspond to the experiment in the DAC (piston cylinder cell). $T_C$ vanishes at $p_c\approx 0.9 $~GPa in both experiments. The superconducting transition $T_{sc}$ is determined by the midpoint of the transition. The inset shows the pressure dependence of the width of the superconducting transition,  $\Delta T_{sc}$.} 
\label{phasediagram}
\end{center}
\end{figure}

Figure \ref{phasediagram} (a) shows the resistivity $\rho$ of UCoGe versus temperature $T$ for various pressures $p$ in the DAC and (b) the $(p,T)$ phase diagram extracted from these measurements which is in good agreement with previous reports \citep{Hassinger2008b, Slooten2009, Hassinger2010a, Bastien2015}. At the lowest pressures the resistivity shows a kink at $T_C$, as indicated in Fig.~\ref{phasediagram}(a) for $p=0.4$~GPa. $T_C$ decreases with pressure from $T_C\approx 2.5~$K at $p=0$~GPa and coincides with the superconducting transition temperature at $p\approx 0.8~$GPa and $T_{sc}=0.8$~K. The extrapolation of the pressure dependence of $T_C$ down to zero temperature determines $p_c\approx 0.9~$~GPa. The exact $p$ dependence of $T_C$ in the SC dome is still unclear \cite{Slooten2009, Hassinger2010a} but from symmetry arguments the transition line should be first order \cite{Mineev2008}. For the second sample measured in the piston cylinder cell a similar phase diagram has been determined [see Fig.~\ref{phasediagram}(b)] with slightly higher $T_C$ and $T_{sc}$. Compared to previous reports  \cite{Slooten2009}\cite{Hassinger2010a}\cite{Bastien2015} the value of $T_C$ and also the exact location of $p_c$ appears strongly sample dependent. However, each study shows that $T_{sc}$ is maximum at or slightly below the pressure where it reaches $T_C$. The inset in Fig.~\ref{phasediagram}(b) shows the transition width $\Delta T_{sc}$ as a function of pressure.  At the maximum of $T_{sc}$ the superconducting transition is sharpest ($\Delta T_{sc} = 40$~mK) and broadens with increasing pressure in the PM phase. $T_{sc}$ is maximal at 0.65~GPa, below $p_c$. Zero resistivity can be observed up to 3.5$~$GPa. Taking the midpoint or the onset of the transition criteria SC vanishes around 4$~$GPa or 4.5$~$GPa, respectively. Thus, SC survives in the PM regime far above $p_c$. The broadening of the superconducting transition under pressure is independent of sample  quality and high pressure conditions and it is clearly related with the strength of $\partial T_{sc}/\partial p$. 

\begin{figure}[!t]
\begin{center}
\includegraphics[width=0.9\linewidth]{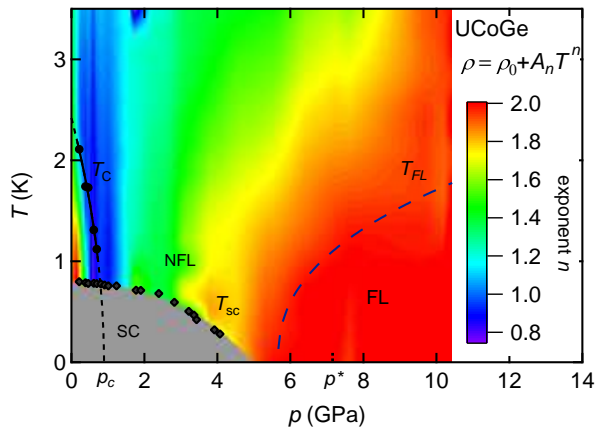}
\caption{(Color online) Colorplot of the resistivity exponent $n$ as a function of temperature and pressure from fitting $\rho = \rho_0 + A_nT^n$ over a sliding window of 300~mK.
$T_C$ and the onset of the superconducting transition as a function of pressure are represented by solid circles and diamonds, respectively. Linear resistivity is observed around $p_c$. At high pressure Fermi-liquid behavior is recovered and the upper limit of the Fermi-liquid regime $T_{FL}$ is indicated by the dashed line.} 
\label{color}
\end{center}
\end{figure}

The $T$-dependence of the resistivity  has been parameterized by fitting a power law $\rho (T)= \rho_0 + A_nT^n$ in the normal state. Here $\rho_0$ is the residual resistivity and $A_n$ gives the strength of the temperature dependent scattering term. In the case of a Fermi-liquid $n=2$ and the coefficient $A_2$ is related to the average effective mass $m^\star$ of the quasiparticles. In multiband heavy-fermion materials  $A_2 \propto (m^\star)^2$ is generally obeyed. To evaluate the pressure and temperature dependence of the exponent $n$ we performed fits with the power law on an sliding window of 0.3~K in the normal state. This allows to plot $n(p,T)$ in Fig.~\ref{color}. The resistivity follows a $T^2$ behavior at ambient pressure and low temperature in the ferromagnetic state. Remarkably, $\rho (T)$ is linear  around $p_c$ above the superconducting transition $T_{sc}$. The $T^2$ behavior of the Fermi-liquid regime is recovered only above 5$~$GPa in the PM state far above $p_c$ below $T_{FL}$.  

The unusual quasi-linear $T$ dependence of $\rho (T)$ observed close to the ferromagnetic instability is not expected in the  spin-fluctuation theory. Theoretically, a $T^{5/3}$ temperature dependence is predicted in the quantum critical regime for a three dimensional ferromagnet \cite{Mathon1968, Moriya1985, Millis1993, Lonzarich1997, Moriya2003}. This is a consequence of an underlying quasiparticle scattering rate that varies linearly with the excitation energy $E$ of a quasiparticle near the Fermi level. 
A $T^{5/3}$ behavior was reported in several itinerant ferromagnetic systems like e.g.~NiAl$_3$\cite{Niklowitz2005}, URhAl\cite{Shimizu2015} and U$_3$P$_4$\cite{Araki2015}. However, strong deviations from this $T^{5/3}$ have been reported for MnSi \cite{Pfleiderer2001, Pedrazzini2006} and ZrZn$_2$ \cite{Smith2008, Kabeya2012} where a $T^{3/2}$ dependence has been observed up to $3p_c$.  

\begin{figure}[!t]
\begin{center}
\includegraphics[width=1\linewidth]{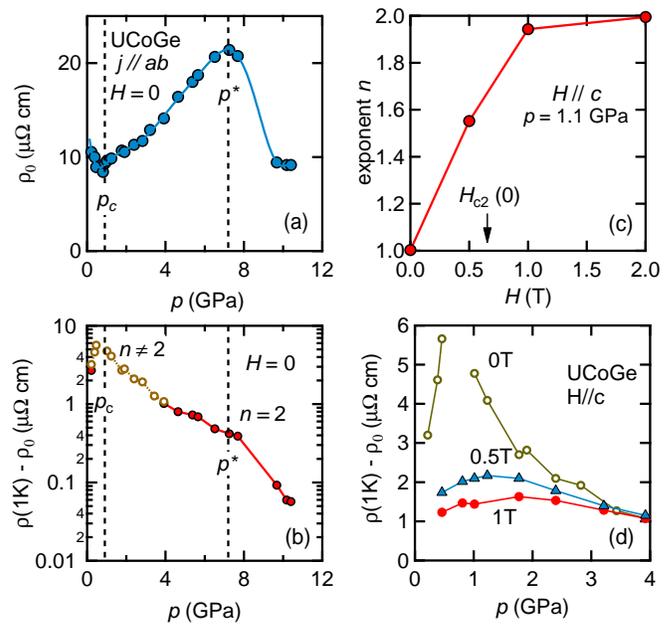}
\caption{(Color online) (a) Residual resistivity $\rho_0$ as a function of pressure. $\rho_0$ shows a minimum at $p_c\approx 0.9~$GPa and a maximum at $p^\star=7.2$~GPa. (b) $\rho(1{\rm K}) - \rho_0$ on a logarithmic scale as a function of pressure. (c) Field dependence of $n$ slightly above $p_c$ obtained from fits in the normal state for $T < 1.5$~K. (d) $\rho(1{\rm K}) - \rho_0$ for different magnetic fields as a function of pressure. } 
\label{r0A}
\end{center}
\end{figure}

A $T$-linear resistance has been observed in different strongly correlated electron systems, like high-$T_c$ cuprate or iron-pnictide superconductors close to the optimal doping  \cite{Lee2006, Daou2009, Cooper2009}, in organic superconductors \cite{Doiron-Leyraud2009}, in ruthenates\cite{Bruin2013}, and  also in several heavy fermion systems, when these are close to quantum criticality such as CeCoIn$_5$ \cite{Sidorov2002}, CeRhIn$_5$\cite{Knebel2008, Park2008}, or YbRh$_2$Si$_2$  \cite{Gegenwart2002}. For the AF heavy fermions, different theoretical scenarii have emerged to explain the unusual $T$ dependence, such as a reduced dimensionality of the magnetic fluctuations \cite{Millis1993, Moriya2003, Rosch1997}, critical valence fluctuations \cite{Holmes2004, Miyake2014}, or fluctuations associated with the change of the electronic structure from the ordered to the PM state  \cite{Senthil2008, Pfau2012, Paul2013}. The specific case of ferromagnetic fluctuations remains to be treated. 

The residual resistivity $\rho_0$ obtained from the fit is represented in Fig.~\ref{r0A}(a) as a function of pressure.  $\rho_0 (p)$ shows shallow a minimum at $p_c\approx0.9$~GPa. It increases up to $\rho_0 \approx 22 \;\mu\Omega$cm at $p^\star \approx 7.2$~GPa and decreases strongly with pressure. Finally it saturates around 9.5~GPa at $\rho_0 = 10 \;\mu\Omega$cm.  $p^\star$ is independent of magnetic field at least up to 7~T. 
The low temperature inelastic electronic scattering at zero magnetic field as a function of pressure is represented on Fig.~\ref{r0A}(b). $\rho (1{\rm K}) - \rho_0$ shows a clear maximum at $p_c$. The decrease of $\rho (1{\rm K}) - \rho_0$ with pressure gets stronger above $p^\star=7.2~$GPa. 
The compressibility of UCoGe was computed by DFT \cite{Yu2011} and experimentally determined by an x-ray scattering experiment under hydrostatic pressure up to 30$~$GPa\cite{Adamska2010}. 
At $p=10$~GPa the volume of the unit cell  is reduced by 3\%. 
No structural transition was observed in the x-ray scattering and only tiny anomalies in the lattice parameters as a function of pressure at $p^\star$ \cite{Adamska2010}.  In the ferromagnetic state at $p=0$ the valence of UCoGe is close to the U$^{3+}$ configuration \cite{Fujimori2012}. It was estimated at 3.2 by LDA calculations for ambient pressure\cite{Samsel-Czekala2010} is expected to be closer to the U$^{4+}$ under pressure.  Thus the anomaly at $p^\star$ may be related to a weak valence crossover as observed in various Ce or Yb based heavy fermion systems under pressure \cite{Holmes2004, Rueff2011, Miyake2014}.

For a small magnetic field of 1~T along the $c$ axis, slightly higher than the upper critical field $H_{c2}$ , the exponent $n=2$ is recovered [see Fig.~\ref{r0A}(c)] and  Fermi-liquid behavior in all the pressure range. The  low temperature electronic scattering, determined by  $\rho (1{\rm K}) - \rho_0$, is plotted  for different magnetic fields as a function of  pressure in Fig.~\ref{r0A}(d). Its acute enhancement at $p_c$ is suppressed under magnetic field and a rather smooth pressure dependence is achieved. This indicates that the low field behavior of UCoGe is determined by magnetic fluctuations which are rapidly suppressed by magnetic field applied along the $c$ axis as has been shown at zero pressure by NMR \cite{Hattori2012}, thermal transport \cite{Taupin2014}, and specific heat experiments \cite{Aoki2011, Wu2016}.

\begin{figure}[!t]
\begin{center}
\includegraphics[width=0.9\linewidth]{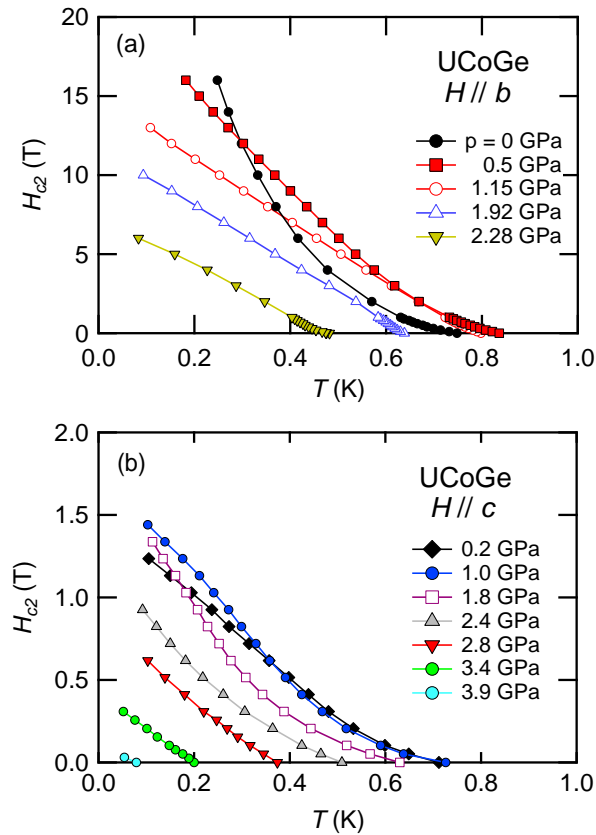}
\caption{(Color online) (a) Temperature dependence of upper critical field $H_{c2}$ in UCoGe for different pressures for field along $b$ axis. Measurements were performed on sample 2 in a piston cylinder cell. The midpoint of resistivity drop was chosen as criteria for the transition. (b) Temperature dependence of $H_{c2}$ in UCoGe for different pressures for field along $c$ axis. Measurements were performed on sample 1 in the DAC. }
\label{Hc2}
\end{center}
\end{figure}

A striking property of UCoGe is its unusual upper critical field $H_{c2}$ with the very strong anisotropy between $H_{c2}$ for field along the $a$ ($H_{c2}^a$) or $b$ ($H_{c2}^b$) hard axis and $H_{c2}$ along the easy magnetization $c$ axis ($H_{c2}^c$). While $H_{c2}^a$ and $H_{c2}^b$ exceed values above 20~T at ambient pressure, $H_{c2}^c$  is as small as 1~T or even less, depending on the sample quality \cite{Huy2008, Aoki2009, Bastien2015}. In addition the temperature dependence $H_{c2} (T)$ for all directions  is very unusual with a strong upward curvature. 
 
$H_{c2}^b (T)$ for field applied along the $b$ axis is represented in Fig.~\ref{Hc2}(a) for different pressures. At ambient pressure $H_{c2}^b (T)$ shows a strong upward curvature with decreasing temperature. This measurement does not show the "S" shape of $H_{c2}$ observed in \cite{Aoki2009}. It may be due to a small misalignment along the $a$ axis of the sample inside the pressure cell.
The unusual behavior of $H_{c2}^b$ along the hard axis is related to the field decrease of $T_C$ in the transverse magnetic field and the concomitant  reinforcement of the magnetic fluctuations \cite{Aoki2009, Aoki2014a}.  This mechanism is quite similar to that of the field reentrance of SC in URhGe \cite{Miyake2008, Mineev2014}. 
 The upward curvature of $H_{c2}^b$ is reduced under pressure and vanishes around $p\approx~1$~GPa. Up to the highest pressure $p=2.28$~GPa, $H_{c2}^b(T)$ is still nearly linear in $T$. $H_{c2}^b (0)$ decreases almost linear with pressure from about $H_{c2}^b(0) \approx 25$~T at zero pressure to 7~T at 2.28~GPa.  The upward curvature disappears when ferromagnetism collapses.\cite{commentAdV}

For $H\parallel c$, the field and pressure evolution of the inelastic low temperature electronic scattering $\rho(1{\rm K}) - \rho_0$ shows that magnetic fluctuations are strongly suppressed with field. The relatively low value of $H_{c2}^c$ and the unusual upward curvature of $H_{c2}^c$ was explained by several models including a magnetic field dependent pairing interaction~\cite{Hattori2012, Tada2013, Mineev2010, Wu2016}. In our experiment the unusual curvature of $H_{c2}^c (T)$ was observed in the whole pressure range of the superconducting dome. 
Inside the framework of equal spin triplet pairing, $H_{c2}(0)$ is governed by the orbital effect and so is proportional to $(m^\star T_{sc})^2$. Here we have to account for a field dependence ($m^\star_H$) which reflects a field dependence of the pairing interaction. The value $T_{sc}(m^\star_H)$ at finite field is lower than  $T_{sc}(m^\star_0)$ expected from zero magnetic field and this leads to the initial curvature of $H_{c2}(T)$. 
However, at $3.4~$GPa far away from the quantum critical region, the ferromagnetic fluctuations have been strongly suppressed and their field dependence become weak. 
Therefore,  $H _{c2}$ at $3.4~$GPa $H_{c2}^c(T)$ varies linearly with temperature and is much closer to the conventional  orbital limitation with a field independent pairing than $H_{c2}$ in the vicinity of $p_c$.  This suggests, that $H_{c2}^c(T)$ is driven by the suppression of pairing interaction under field. 

\section{Conclusion}

To summarize we determined the $(p,T)$ phase diagram of UCoGe up to 10.5$~$GPa by resistivity measurements. $T_C$ vanishes at $p_c\approx 0.9~$GPa and SC has been observed up to $4 p_c$. Ferromagnetic fluctuations account for the unusual normal state transport properties observed up to 5~GPa  as well as the pressure and field dependence of $H_{c2}$.  
The superconducting critical field $H_{c2}^c$ for field along the $c$ axis is determined by the pressure and field dependence of the ferromagnetic fluctuations which are suppressed for $H\approx 1$~T at $p=0$. $H_{c2}^b$ for transverse magnetic field changes drastically on crossing the ferromagnetic phase boundary below $p_c$. Entering in the paramagnetic regime above $p_c$ $H_{c2}^b$ recovers rapidly the conventional initially linear $T$ dependence of the upper critical field. Finally, the crossover pressure $p^\star$ marks a change in the magnetic and electronic properties linked to the pressure dependence of the 5$f$ occupation number and a smooth crossover from a strongly to weakly correlated regime.

Our experiment gives now a sound basis on the anisotropic SC response of $H_{c2}$ on the static and dynamic components of the ferromagnetism. To our knowledge, it is an unique example where the magnetic coupling is deeply modified (here for $H \parallel c$) in low magnetic fields. Close to $p_c$ the comparable strength of  $T_C$ and $T_{sc}$ leads to remarkable non-Fermi-liquid behavior which indicates the strong coupling of magnetism and superconductivity. This is remarkably different from other ferromagnetic superconductors driven to quantum criticality by pressure (UGe$_2$)\cite{Tateiwa2001, Terashima2006} or magnetic field (URhGe)\cite{Gourgout2016} where a Fermi-liquid regime is robust.

\begin{acknowledgements}
We are grateful to J.-P. Brison, B. Wu, A. Pourret, K. Ishida, V. Mineev, and A.~de Visser for insightful discussions. 
We acknowledge support from ERC grant "NewHeavyFermion'', the French ANR "PRINCESS'', and KAKENHI (25247055,15H05884,15H05882, 15K21732, 16H04006). 
\end{acknowledgements}


\begin{thebibliography}{75}
\bibitem [{\citenamefont {Wu}\ and\ \citenamefont {et~al.}()}]{Wu2016}%
  \BibitemOpen
  \bibfield  {author} {\bibinfo {author} {\bibfnamefont {B.}~\bibnamefont
  {Wu}}\ and\ \bibinfo {author} {\bibnamefont {et~al.}},\ }\href@noop {}
  {}\bibinfo {note} {To be published}\BibitemShut {NoStop}%
\end{thebibliography}

%

\end{document}